\documentclass[onecolumn]{emulateapj}  
\usepackage{epstopdf}
\usepackage{ulem}
\usepackage{amssymb}
\usepackage{natbib}
\usepackage{times}
\usepackage{graphicx}
\usepackage[usenames,dvipsnames]{pstricks}
 \usepackage{psfrag}

\def\aj{AJ}
\def\araa{ARA\&A}
\def\apj{ApJ}
\def\apss{Ap\&SS}
\def\aap{A\&A}
\def\jcap{J. Cosmology Astropart. Phys.}
\def\mnras{MNRAS}
\def\prd{Phys.~Rev.~D}
\def\physrep{Phys.~Rep.}

\newcommand{\be}{\begin{equation}}
\newcommand{\ee}{\end{equation}}
\newcommand{\bary}{\begin{eqnarray}}
\newcommand{\eary}{\end{eqnarray}}
\newcommand{\en}{E_\nu}

\shorttitle{}
\shortauthors{}
\begin{document}
\title{The hadronic picture of the radiogalaxy M87}
\author{A. Marinelli$\dagger$\altaffilmark{1}, N. Fraija$\dagger\dagger$\altaffilmark{2} and B. Patricelli\altaffilmark{2}} 
\email{antonio.marinelli@fisica.unam.mx, nifraija@astro.unam.mx and bpatricelli@astro.unam.mx} 
\altaffiltext{1}{Instituto de F\'isica, Universidad Nacional Aut\'onoma de M\'exico, Circuito Exterior, C.U., A. Postal 70-264, 04510 M\'exico D.F., M\'exico.}
\altaffiltext{2}{Instituto de Astronom\' ia, Universidad Nacional Aut\'onoma de M\'exico, Circuito Exterior, C.U., A. Postal 70-264, 04510 M\'exico D.F., M\'exico.}
\altaffiltext{$\dagger$}{Actually at I.N.F.N. and Physics Institute of Pisa University, Edificio C - Polo Fibonacci Largo B. Pontecorvo, 3 - 56127 Pisa, Italy.}
\altaffiltext{$\dagger\dagger$}{Luc Binette-Fundaci\'on UNAM Fellow.}
%
%
\begin{abstract}
Very high energy gamma-ray emission of Fanaroff-Riley I objects is not univocally explained by a single emission model. Leptonic models with one and multi-zone emission regions, occurring 
in the jet of these objects, are usually used to describe the broadband spectral energy distribution. A correlation between the X-ray and TeV emission is naturally expected within 
leptonic models whereas a lack of correlation between these two observables represents a challenge and favors the hadronic scenarios. This is the case of M87 as we show here by analyzing its TeV 
and X-ray emission recorded in the last decade. Furthermore, we point out that the spectra obtained by MAGIC, H.E.S.S. and VERITAS telescopes cannot be described with the same leptonic model 
introduced by the Fermi-LAT collaboration. We introduce hadronic scenarios to explain the TeV gamma-ray fluxes of this radiogalaxy as products of Fermi-accelarated protons interacting with seed 
photons in the jet or thermal particles in the giant lobes. By fitting this part of spectral energy distribution as pion decay products, we obtain the expected neutrino counterpart and the luminosity of 
accelerating protons in the jet and/or lobes. With the expected neutrino fluxes we investigate, through Monte Carlo simulations, the possibility to see the signal from M87 with a Km$^{3}$ neutrino 
telescope, and compare the results with what has been seen by IceCube experiment up to now. Finally we constrain the features of giant lobes through the observations performed at ultra high 
energies by TA experiment.
\end{abstract}

\section{Introduction}
The nearby radio galaxy M87 is located in the Virgo galaxy cluster at a distance of $\sim$ 16 Mpc (z=0.0043) and hosts a central black hole of (3.2 $\pm$0.9) $\times$ 10$^9$ solar masses 
\citep{2007ApJ...655..144M}.  M87, being one of the nearest radiogalaxies, is among the best-studied of its class. It has been detected at energy ranges from radio to very high 
energy (VHE) gamma-rays \citep{2012ApJ...746..151A}. The radio emission of M87 comes from the inner lobes, the intermediate ridges and the outer diffuse ``halo'', respectively at distances of  $\sim
$2.5 kpc (30"), $\sim$15 kpc (3") and $\sim$40 kpc (8") from the nucleus \citep{2002ApJ...579..560Y}. The gamma-ray emission of M87 has been detected by Large Area Telescope (LAT) at MeV-
GeV energy ranges \citep{2009ApJ...707...55A}. The entire spectral energy distribution (SED), including the radio, the X-ray \citep{spa96, tan08}, the Chandra and VLBA measurements \citep{bir91, 
des96} and the LAT gamma-ray data \citep{2009ApJ...707...55A}, is well fitted with a single-zone Self Synchrotron Compton (SSC) emission model as showed by the Fermi-LAT collaboration 
\citep{2009ApJ...707...55A}. In this work we extend the analysis of the SED up to higher energies considering also the data collected by H.E.S.S., MAGIC and VERITAS \citep{2012ApJ...746..151A, 
2010ApJ...716..819A, 2006Sci...314.1424A} between 2004 and 2010. The TeV gamma-ray data obtained from these Imaging Atmospheric Cherenkov Telescopes (IACTs) data cannot be explained 
by the same SSC scenario so that additional mechanisms should be considered to explain the whole spectrum (for example hadronic processes; \citep{2012grb..confE.131F} or multi-zone SSC). 
Here we investigate two different hadronic scenarios to describe the TeV gamma-ray data: the interaction of accelerated protons in the jet of M87 with the MeV SSC photons and the interaction of 
boosted protons in the giant lobes \citep{2014ApJ...783...44F} with the thermal protons target. The individual SEDs obtained with H.E.S.S., MAGIC and VERITAS campaigns have been well fitted 
considering the proton-proton (pp) and the proton-gamma (p$\gamma$) scenarios with different sets of spectral parameters. We also compared simultaneous activities in gamma-rays (with the above 
mentioned data) and in X-rays, with data from RXTE-ASM \citep{2006AdSpR..38.2959S} on different time scales. The emission in these two energy bands resulted not significantly correlated 
providing further hints of hadronic origin of the TeV gamma-rays observed. Given these results, we consider the possibility of having a neutrino counterpart emitted by M87 and we obtain the expected 
neutrino spectra from the best fit gamma-ray spectra of the three different data samples. We introduce the obtained neutrino spectra in a Monte Carlo simulation of a hypothetical Km$^{3}$ neutrino 
telescope in the north hemisphere and we get the signal to noise ratio for one year of data-taking. Since the observation of M87 with a neutrino telescope cannot  be considered preferential from one 
of the two hemispheres, it has been possible to compare this result with what has been observed up to now by the IceCube experiment. Finally, within the presented hadronic scenario we investigate 
the possibility for protons to be accelerated up to Ultra High Energies  (UHEs) and put limits on the density of M87 giant lobes considering different sizes. For this calculation, we use the accelerated 
proton spectra obtained through the spectral fit of TeV gamma-ray (within the pp interaction scenario). Then, we constrained the properties of the lobes taking into account the UHE proton flux 
observed by Telescope Array (TA) experiment \citep{2012NIMPA.689...87A} in 5 years of data-taking.
\newpage
\section{Hadronic Interactions}
Radiogalaxies have been proposed as powerful accelerators of charged particles through the Fermi acceleration mechanism \citep{2007Ap&SS.309..119R}. The Fermi-accelerated protons can be described by a simple power law
\be\label{prot_esp}
\frac{dN_p}{dE_p}=A_p E_p^{-\alpha}\,,
\ee
where $\alpha$ is the power index and $A_p$ is the proportionality constant.   For this work, we consider that these protons are cooled down  by p$\gamma$ and pp interactions occurring in the jet and giant  lobes, respectively.  Both interactions produce VHE gamma-rays and neutrinos as explained in the following subsections. We hereafter use primes (unprimes) to define the quantities in a comoving (observer) frame,  c=$\hbar$=1 in natural units and redshift z$\simeq$ 0.
\subsection{p$\gamma$ interaction}
Charged ($\pi^+$) and neutral ($\pi^0$) pions are obtained from p$\gamma$ interaction through the following channels   
 \begin{eqnarray}
p\, \gamma &\longrightarrow&
\Delta^{+}\longrightarrow
\left\{
\begin{array}{lll}
p\,\pi^{0}\   &&   \mbox{fraction }2/3, \\
n\,  \pi^{+}      &&   \mbox{fraction }1/3,\nonumber
\end{array}\right. \\
\end{eqnarray}
and neutral pions decay into photons, $\pi^0\rightarrow \gamma\gamma$,  carrying $20\%\,(\xi_{\pi^0}=0.2)$ of the proton's energy $E_p$.  As has been pointed out by Waxman and Bahcall  \citep{PhysRevLett.78.2292}, the photo-pion spectrum is obtained from the efficiency of this process 
\begin{equation}
f_{\pi^0,p\gamma} \simeq \frac {t'_{dyn}} {t'_{\pi^0}}  =\frac{r_d}{2\,\delta_{D}\,\gamma^2_p}\int\,d\epsilon\,\sigma_\pi(\epsilon)\,\xi_{\pi^0}\,\epsilon\int dx\, x^{-2}\, \frac{dn_\gamma}{d\epsilon_\gamma} (\epsilon_\gamma=x)\,,
\end{equation}
where $t'_{dyn}$ and $t'_{\pi^0}$ are the dynamical and the pion cooling times, $\gamma_p$ is the proton Lorentz factor, $r_{d}=\delta_{D}dt$ is the comoving dissipation radius as function of Doppler factor ($\delta_{D}$) and observational time ($t^{obs}$), $dn_\gamma/d\epsilon_\gamma$ is the spectrum of target photons,  $\sigma_\pi(\epsilon_\gamma)=\sigma_{peak}\approx 9\times\,10^{-28}$ cm$^2$ is the cross section of pion production. Solving the integrals we obtain 
{\small
\bary
f_{\pi^0,p\gamma} \simeq \frac{L_\gamma\,\sigma_{peak}\,\Delta\epsilon_{peak}\,\xi_{\pi^0}}{8\pi\,\delta_D^2\,r_d\,\epsilon_{\gamma,b}\,\epsilon_{peak}}
\cases{
\left(\frac{\epsilon_{\pi^0,\gamma,c}}{\epsilon_{0}}\right)^{-1} \left(\frac{\epsilon_{\pi^0,\gamma}}{\epsilon_{0}}\right)       &  $\epsilon_{\pi^0,\gamma} < \epsilon_{\pi^0,\gamma,c}$\cr
1                                                                                                                                                                                                            &   $\epsilon_{\pi^0,\gamma,c} < \epsilon_{\pi^0,\gamma}$\,,\cr
}
\eary
}
where $\Delta\epsilon_{peak}$=0.2 GeV,  $\epsilon_{peak}\simeq$ 0.3 GeV, $L_\gamma$ is the luminosity and $\epsilon_{\gamma,b}$ is the break energy of the seed photon field \citep{2014MNRAS.441.1209F}.  By considering the simple power law for a proton distribution (eq. \ref{prot_esp}) and the conservation of the photo pion flux for this process, $f_{\pi^0,p\gamma}\,E_p\,
(dN/dE)_p\,dE_p=\epsilon_{\pi^0,\gamma}\,(dN/d\epsilon)_{\pi^0,\gamma}\,d\epsilon_{\pi^0,\gamma}$, the photo-pion spectrum is given by
{\small
\bary
\label{pgammam}
\left(\epsilon^2\,\frac{dN}{d\epsilon}\right)_{\pi^0,\gamma}= A_{p\gamma,\gamma}  \cases{
\left(\frac{\epsilon_{\pi^0,\gamma,c}}{\epsilon_{0}}\right)^{-1} \left(\frac{\epsilon_{\pi^0,\gamma}}{\epsilon_{0}}\right)^{-\alpha+3}          &  $ \epsilon_{\pi^0,\gamma} < \epsilon_{\pi^0,\gamma,c}$\cr
\left(\frac{\epsilon_{\pi^0,\gamma}}{\epsilon_{0}}\right)^{-\alpha+2}                                                                                        &   $\epsilon_{\pi^0,\gamma,c} < \epsilon_{\pi^0,\gamma}$\,,\cr
}
\eary
}
\noindent with the normalization energy $\epsilon_0$, proportionality constant of p$\gamma$ interaction given by
\be\label{Apg}
A_{p\gamma,\gamma}= \frac{L_\gamma\,\epsilon^2_0\,\sigma_{peak}\,\Delta\epsilon_{peak}\left(\frac{2}{\xi_{\pi^0}}\right)^{1-\alpha}}{4\pi\,\delta_D^2\,r_d\,\epsilon_{\gamma,b}\,\epsilon_{peak}} \,A_p\,,
\ee
\noindent and the break photon-pion energy given by 
 \be
\epsilon_{\pi^0,\gamma,c}\simeq 31.87\,{\rm GeV}\, \delta_D^2\, \left(\frac{\epsilon_{\gamma,b}}{ {\rm MeV}}\right)^{-1}\,.
\label{pgamma}
\ee
The eq. \ref{pgammam} describes the contribution of photo-pion emission to the SED.

\subsection{pp interaction}
 $\pi^+$ and $\pi^0$ are also obtained from the pp interaction by means of  channel \citep{2008PhR...458..173B,2003ApJ...586...79A,2009herb.book.....D,2002MNRAS.332..215A}
\begin{eqnarray}
p\,+ p &\longrightarrow& \pi^++\pi^-+\pi^0 + X.
\label{pp}
\end{eqnarray}
\noindent Once again neutral pions decay  in two gammas,  $\pi^0\rightarrow \gamma\gamma$,  carrying  33$\%$ ($\xi_{\pi^0}$=0.33) of the proton energy $E_p$. Assuming that accelerated protons interact in the lobe region, spatially constrained by $R$ and thermal particle density $n_p$, we describe the efficiency of the process through
\be
f_{\pi^0,pp}\approx R\,n_p\,k_{pp}\,\sigma_{pp}\,,
\ee
where $\sigma_{pp}\simeq 30(0.95 +0.06\,\rm{ln(E/GeV))}$ mb  is the nuclear interaction cross section and $k_{pp}=1/2$ is the inelasticity coefficient. Taking into account the proton distribution (eq. \ref{prot_esp}) and the conservation of the photo pion flux \citep{2003ApJ...586...79A, 2012ApJ...753...40F, 2002MNRAS.332..215A, 2009MNRAS.393.1041H}
\be\label{fpp}
f_{\pi^0 , pp}(E_p)\,E_p\,\left(\frac{dN_p}{dE_p}\right)^{obs}\,dE_p=\epsilon_{\gamma, {\pi^0}}\,\left(\frac{dN_\gamma}{d\epsilon_\gamma}\right)^{obs}_{\pi^0}\,d\epsilon_{\gamma, {\pi^0}},
\ee
then the observed gamma-ray spectrum can be written as
\begin{equation}
\label{spe_pp}
\left(\epsilon^{2}_\gamma\, \frac{dN_\gamma}{d\epsilon_\gamma}\right)^{obs}_{\pi^0}= A_{pp,\gamma}\, \left(\frac{\epsilon_{\gamma,\pi^0}}{{\rm \epsilon_0}}\right)^{2-\alpha},
\end{equation}
where the proportionality constant of pp interaction is 
\be
A_{pp,\gamma}= R\,n_p\,k_{pp}\,\sigma_{pp}\,(2/\xi_{\pi^0})^{2-\alpha}\,\epsilon_0^2\,A_p\,. 
\label{App}
\ee
The eq. \ref{spe_pp} shows the contribution of pp interactions to  the spectrum of gamma rays produced in the lobes. 
%
%
\section{The VHE neutrino expectation}
The hadronic interactions described above produce also a neutrino counterpart in the jet (trough p$\gamma$) and in the lobes (through pp) of the AGN.
In these processes the neutral pion produced decays into two gammas, $\pi_{0}\rightarrow\gamma\gamma$, and the charged pion into leptons and 
neutrinos, $\pi^{\pm}\rightarrow e^{\pm}+\nu_{\mu}/\bar{\nu}_{\mu}+\bar{\nu}_{\mu}/\nu_{\mu}+\nu_{e}/\bar{\nu}_{e}$. The effect of neutrino oscillations on the 
expected flux balances the number of neutrinos per flavor \citep{2008PhR...458..173B} arriving to Earth. Assuming the described interactions, we expect that the VHE gamma rays 
and the respective neutrino counterpart have a SED strictly linked to the SED of accelerated primary protons.
The spectrum of  neutrino produced by the hadronic interactions can be written as:
\be
\frac{dN_{\nu}}{d\en}=A_{\nu} \, \left(\frac{\en}{\mbox{TeV}}\right)^{-\alpha_{\nu}},
\label{espneu1}
\ee
where the normalization factor, A$_{\nu}$,  is calculated by correlating the neutrino flux luminosity with the TeV photon flux  \citep{2008PhR...458..173B}.  This correlation is given by:
\be
\int \frac{dN_{\nu}}{d\en}\,\en\,d\en=K\int \frac{dN_\gamma}{dE_\gamma}\,E_\gamma\,dE_\gamma\,.
\ee
Where for pp interaction should be used $K=1$ and for p$\gamma$ interaction $K=1/4$ \citep[see, Julia Becker] [and reference therein]{2007Ap&SS.309..407H}. The spectral indices for neutrino and 
gamma-ray spectrum are considered similar  $\alpha\simeq \alpha_\nu$ \citep{2008PhR...458..173B} while the carried energy is slightly different: each neutrino brings 5\%  of the initial proton energy 
($\en=1/20\,E_p$) while  each photon brings around 16.7\%.  With these considerations the normalization factors are related by
\be
A_{(pp,\nu/p\gamma,\nu)}=K\cdot A_{(pp,\gamma/p\gamma,\gamma)}\,\epsilon_0^{-2}\, (2)^{-\alpha+2},
\ee
where A$_{pp,\gamma}$ and A$_{p\gamma,\gamma}$ are given by the Eq. (\ref{App}) and the Eq. (\ref{Apg}) and the factor $2^{-\alpha+2}$ is introduced because the neutrino carries $1/2$ of $\gamma$ energy. 
Therefore extending the spectrum of expected neutrino to maximum energies detectable by a Km$^{3}$ Cherenkov detector array, we can obtain the number expected neutrino events detected as:
\be
N_{ev} \approx\,T \rho_{water/ice}\,N_A\,V_{eff}\,\int_{E_{min}}^{E_{max}}\sigma_{\nu}\,A_{(pp,\nu/p\gamma,\nu)}\left(\frac{E_{\nu}}{TeV}\right)^{-\alpha}dE_{\nu}.
\label{numneu1}
\ee
Where $N_A$ is the Avogadro number, $\rho_{water/ice}$ is the density of environment for the neutrino telescope, $E_{min}$ and $E_{max}$ are the low and high energy threshold considered,
$V_{eff}$ is the $\nu_{\mu}+\bar\nu_{\mu}$ effective volume, obtained through Monte Carlo simulation, for a hypothetical Km$^{3}$ neutrino telescope considering a neutrino source at the declination 
of M87.
\section{UHE cosmic rays phenomenology}
As explained before, for this class of AGN two possible acceleration regions have been identified: one close to black-hole (BH) at a sub-parsec distance and the other one in the extended lobes at 
dozens of kiloparsecs. Relativistic protons could be accelerated up to UHE depending on the properties  of the acceleration region: the magnetic field ($\mathcal B$)  and the size ($\mathcal{R}$). 
Then, the maximum  energy obtained from the source is \citep{1984ARA&A..22..425H}
\be\label{hillas}
E_{max}=Ze\,\mathcal B\,\mathcal R\,\Gamma,
\ee
where the maximum energy reachable is limited in the jet by the emission region ($\mathcal{R}=r_{pc}$) \citep{2009ApJ...707...55A}, and in the lobes by the  magnetic field $\mathcal B=B_{\mu G}$ 
\citep{MagneticFieldAGN}; with $Z$ we indicate the atomic number.\\ Taking into account the values of magnetic field and acceleration region in the jet (B=55 mG and $r_d$=1.4$\times$10$^{16}$ 
cm) \citep{2009ApJ...707...55A} and in the lobes (B=10$^{-6}$ Gauss and R=10 kpc) \citep{MagneticFieldAGN}, the maximum energies achievable $E_{max}$  (eq. \ref{hillas}) are 1.05$
\times10^{19}$ eV and 4.22$\times10^{20}$ eV, respectively.  From these considerations we can claim that UHECRs can reach energies greater that 57 EeV only through the lobe regions. On the 
other hand, UHECRs traveling from source to Earth are randomly deviated by galactic
\be\label{thet_G}
\theta_G\simeq 3.8^{\circ}\left(\frac{E_{p,th}}{57 EeV}\right)^{-1} \int^{L_G}_0  | \frac{dl}{{\rm kpc}}\times \frac{B_G}{4\,{\rm \mu G}} |
\ee
and extragalactic
\be\label{thet_EG}
\theta_{EG}\simeq 4^{\circ}\left(\frac{E_{p,th}}{57 EeV}\right)^{-1} \left( \frac{B_{EG}}{1\,{\rm nG}} \right)\,\left(\frac{L_{EG}}{100\, {\rm Mpc}}\right)^{1/2}\,\left(\frac{l_c}{1\, {\rm Mpc}}\right)^{1/2}   
\ee
magnetic fields.  Here  L$_{EG}$ corresponds to the average path of extragalactic charged particle going through our Galaxy (20 kpc), $l_c$ is the coherence length \citep{1997ApJ...479..290S,
2009JCAP...08..005M} and $E_{p,th}=57\,{\rm EeV}$ is  the threshold energy of the TA experiment.
\subsubsection{Expected Number of  UHECRs} 
TA experiment, located at 1.400 m above sea level in Millard Country, Utah, USA,  is built with three fluorescence detector (FD) stations and a scintillator surface detector (SD) array 
\citep{2012NIMPA.689...87A}. It was designed to observe extensive air showers produced by primary cosmic rays with energies above 1 EeV. This array has an area of $\sim$ 700 Km$^2$ and is in 
data taking since 2008.   To estimate the number of UHECRs,  we consider the TA  exposure for a point source TA\,exp\,=\,$\Xi\,t_{op}\, \omega(\delta_s)/\Omega$, with $\Xi\times\,t_{op}=5\,\rm yr\times7\times10^2\,\rm km^2$. Here $t_{op} $ is the total operational time (from 2008 May 11 and 2013 May 4),  $\omega(\delta_s)$ is an exposure correction factor for the declination of Mrk 421 
\citep{2001APh....14..271S} and $\Omega\simeq\pi$ the experiment covered solid angle. The expected number of UHECRs above an energy $E_{p,th}$ yields
\be
N_{\tiny UHECR}<  ({\rm  TA\,exp})\times \,N_p, 
\label{num}
\ee
where $N_p$ is the  UHECR flux arriving to the detector.  Taking into account that  the proton spectrum is described by eq. (\ref{prot_esp}), then the expected number of UHECRs above a threshold energy $E_{p,th}$ can be written as
\bary
N_{\tiny UHECR} <   \frac{\Xi\,t_{op}\, \omega(\delta_s)}{\Omega\,(\alpha-1)}\,{\rm GeV} \,\left(\frac{E_{p,th}}{\,{\rm GeV}}\right)^{-\alpha+1}\, A_p 
\label{nUHE1}
\eary
with $\alpha$ the spectral index and $A_p$ the proportionality constant. These values can be obtained from the signature of hadronic interactions at lower energies.
%

%
\section{Analysis and results}
\begin{figure}
\centering
\includegraphics[width=0.8\textwidth]{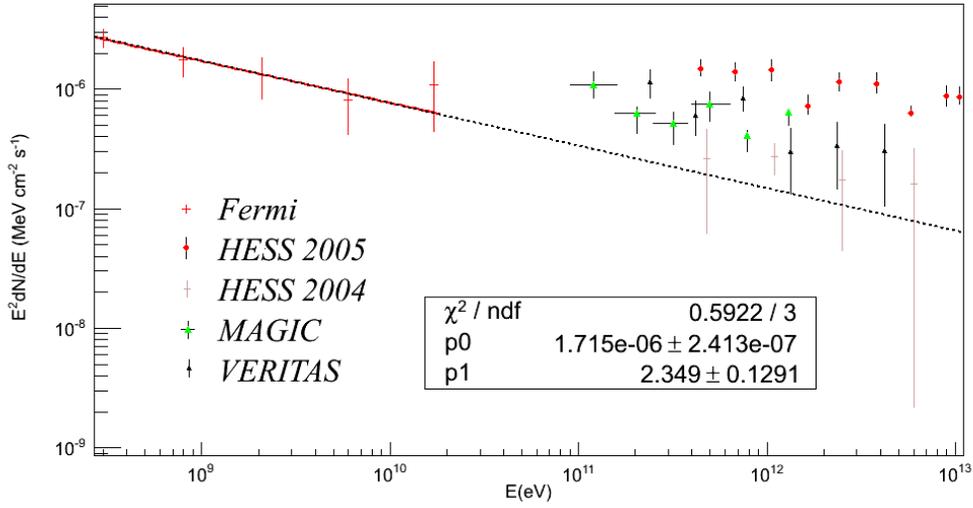}\\
\caption{SED of M87 from 300 MeV up to more than 10 TeV. The parameters of the fit of Fermi data are reported as a $p_{0}$ (proportionality constant) and $p_{1}$ (spectral index). It is evident from this plot that only the one-zone SSC model used by the Fermi collaboration cannot be used to fit at the same time Fermi data and the reported TeV campaigns of MAGIC, VERITAS and H.E.S.S.}\label{fit}
\end{figure}

\begin{figure}
\centering
\begin{tabular}{ccc}
\includegraphics[width=0.3\textwidth]{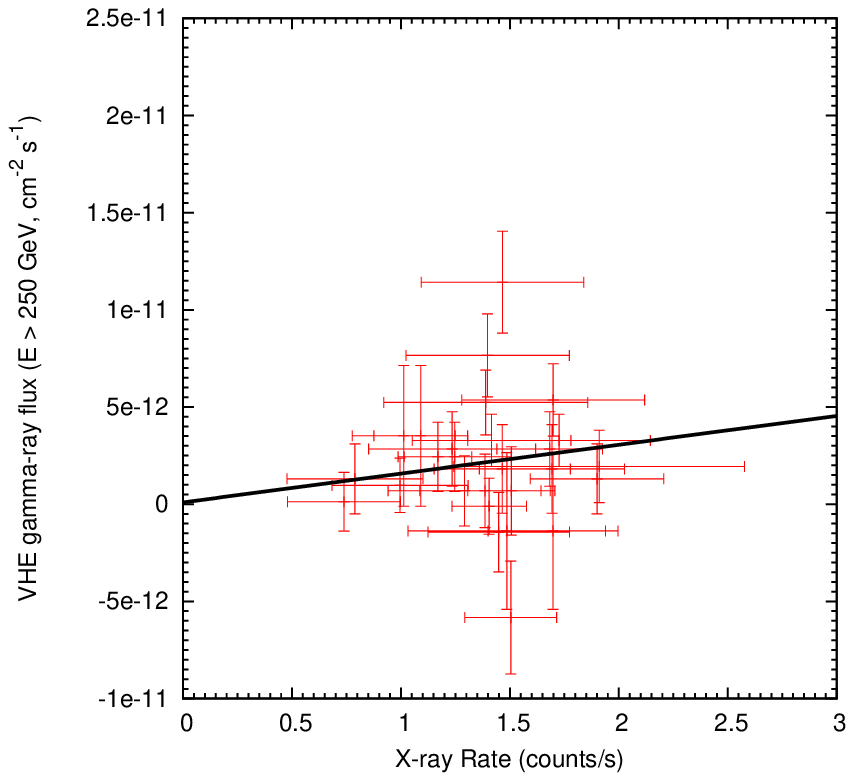}  &
\includegraphics[width=0.3\textwidth]{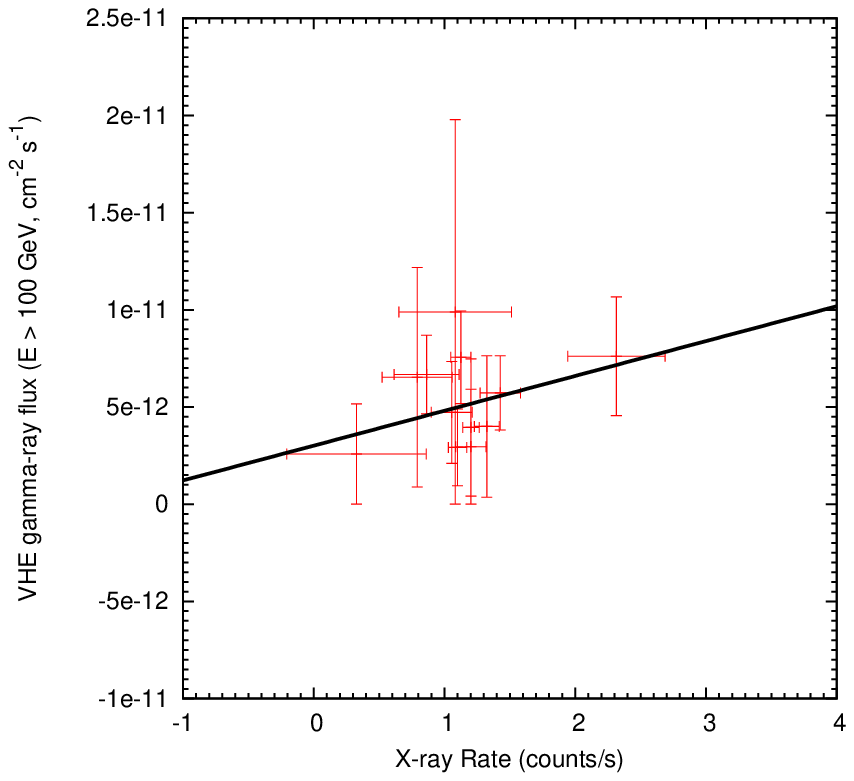}  &
\includegraphics[width=0.3\textwidth]{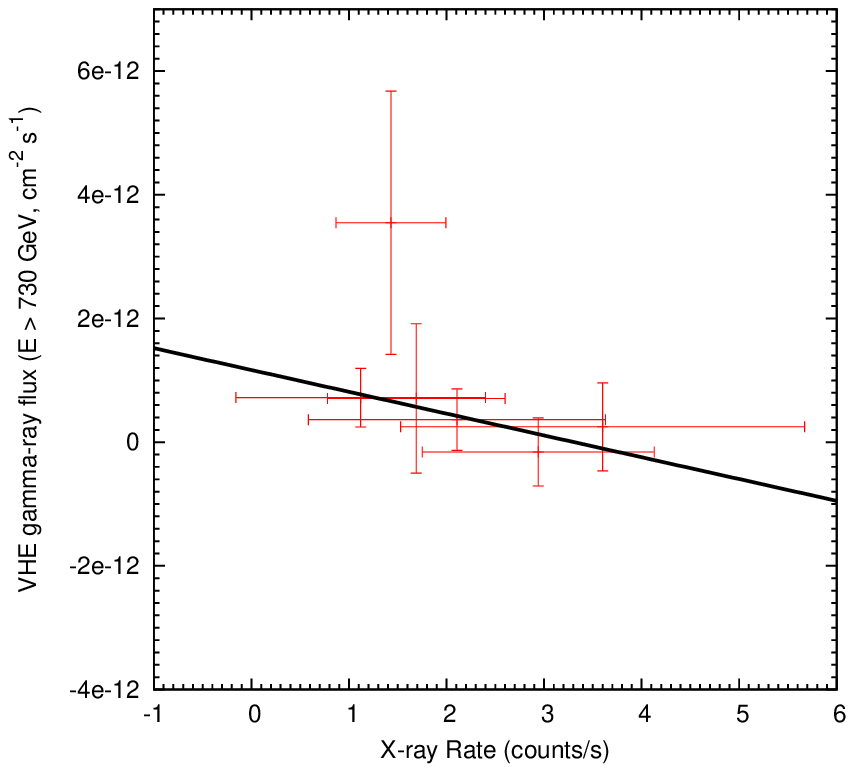}
\end{tabular}
\caption{X-ray/VHE $\gamma$-ray correlation, obtained with X-ray data from RXTE/ASM and $\gamma$-ray data from VERITAS (left panel), MAGIC (central panel) and H.E.S.S (right panel)}\label{event-display}
\end{figure}

\begin{figure}
\centering
\includegraphics[width=0.85\textwidth]{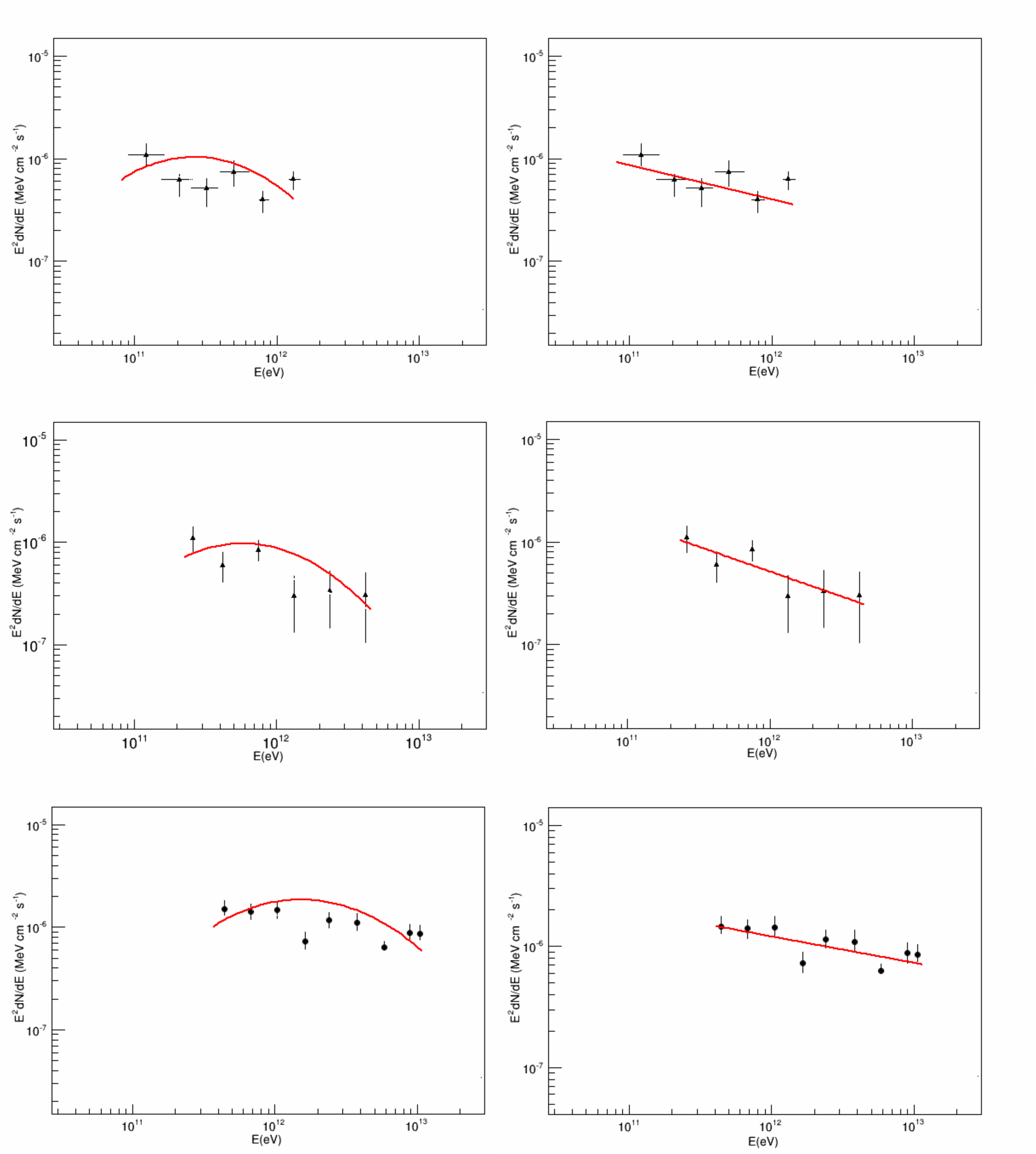}\\
\caption{Best fit of the SED data reported by MAGIC, VERITAS and H.E.S.S. from top to bottom. The left column represents the proton-photon interaction model with the parameters reported in table A1 while the right column represents the proton-proton interaction model with the parameters reported in table A2.}\label{fit2}
\end{figure}

\begin{figure}
\centering
\includegraphics[width=0.8\textwidth]{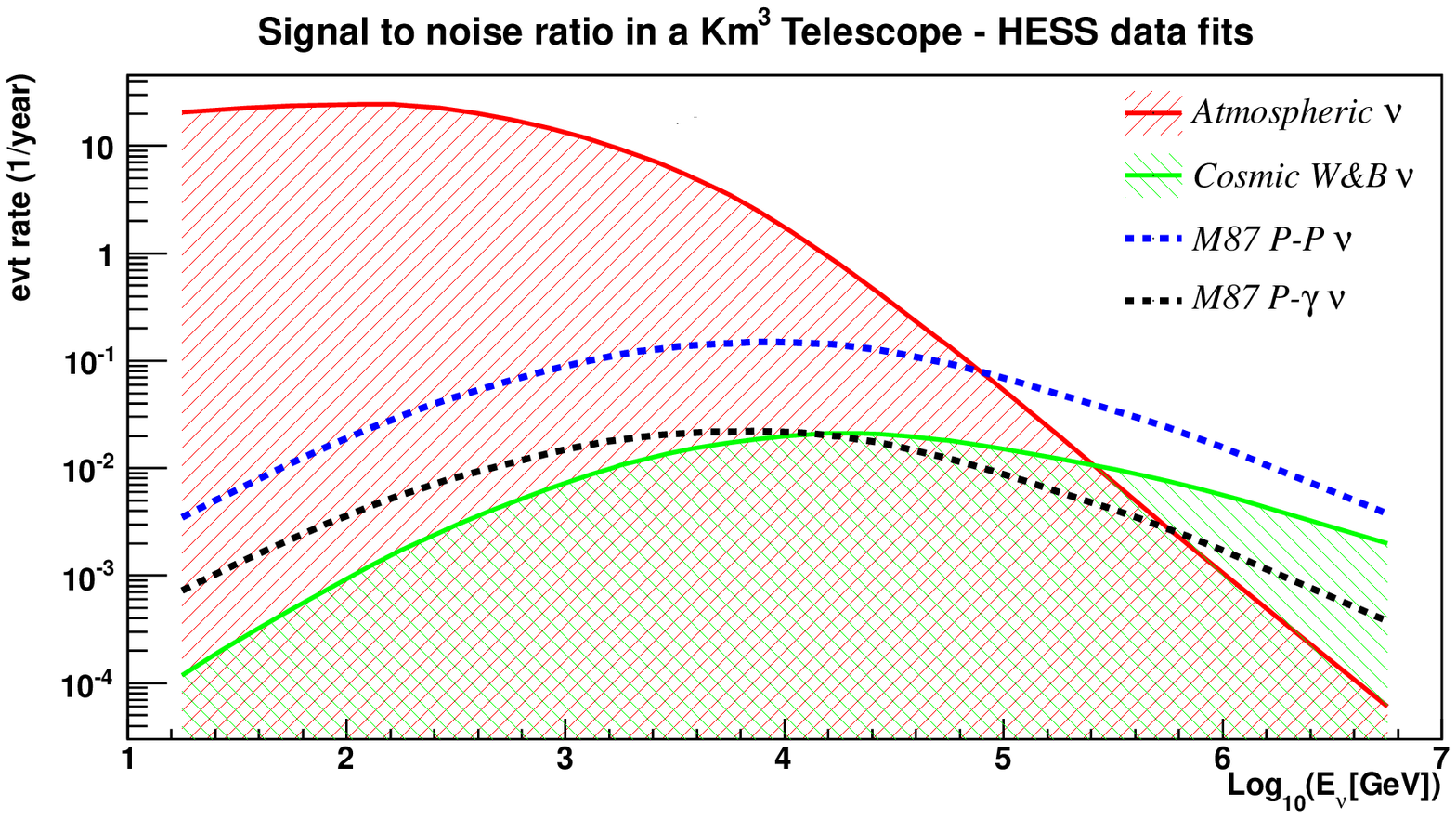}\\
\includegraphics[width=0.8\textwidth]{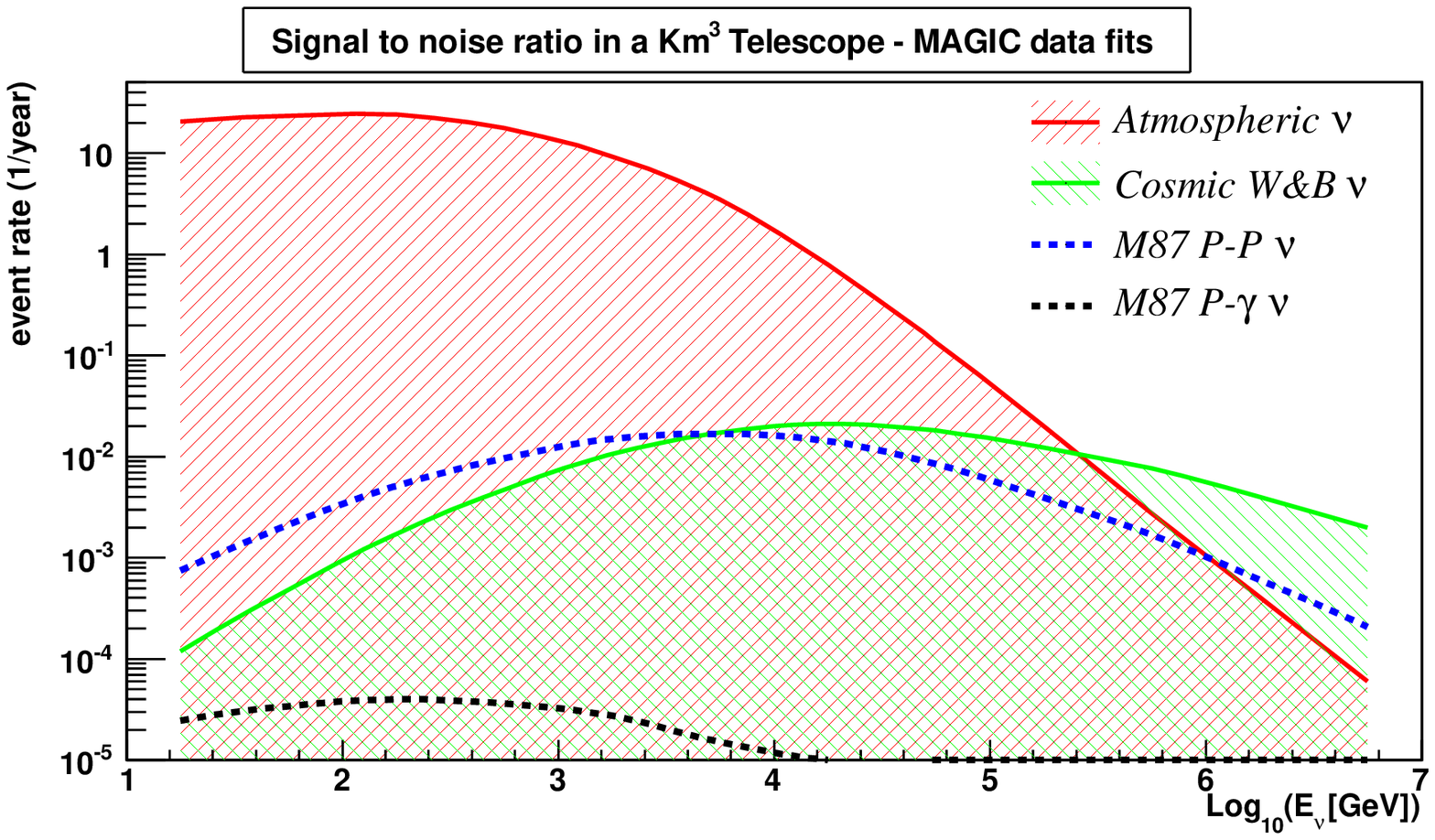}\\
\includegraphics[width=0.8\textwidth]{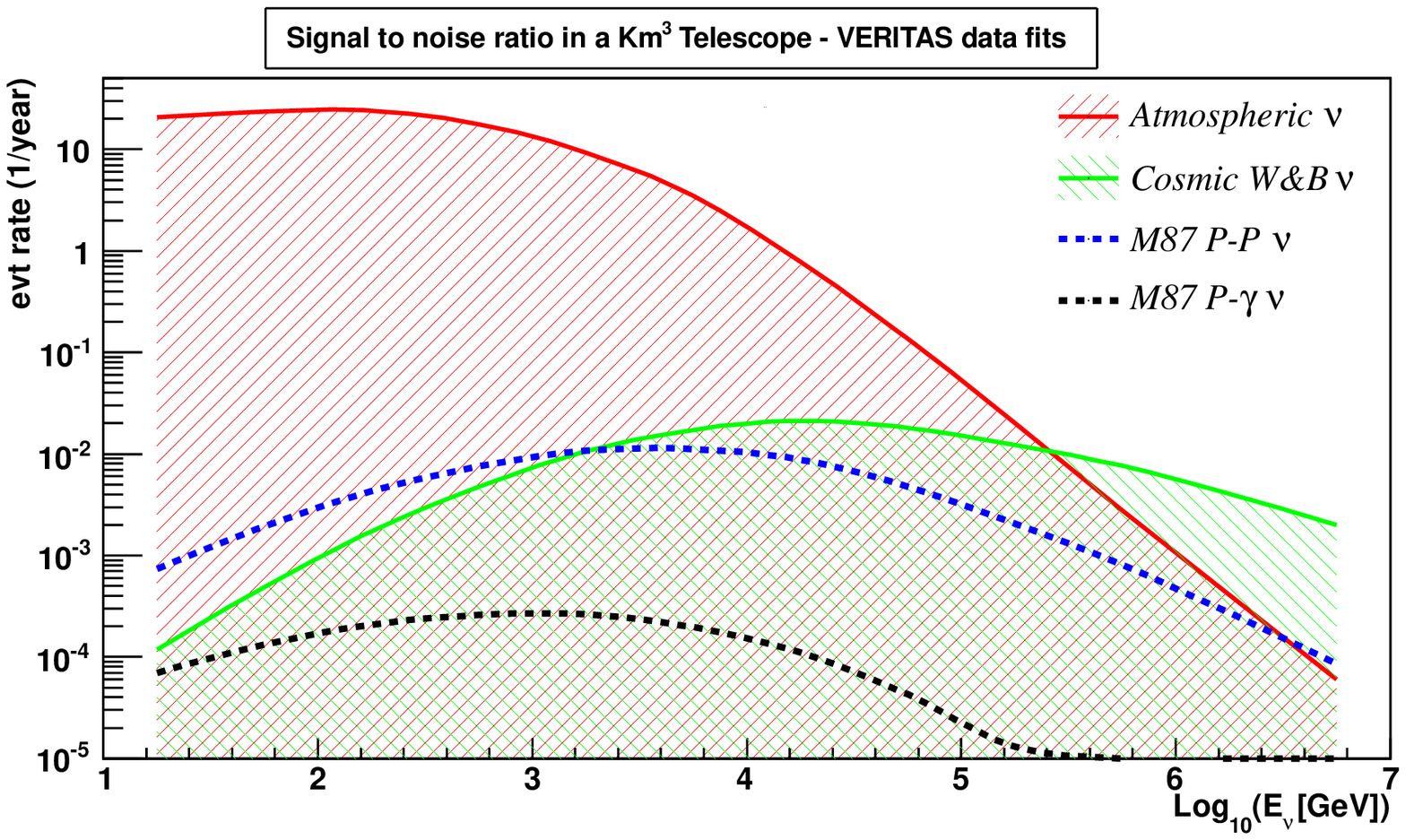}\\
\caption{Neutrino signal to noise ratio for a Km$^{3}$ northern neutrino telescope considering the M87 SEDs reported by H.E.S.S., MAGIC and VERITAS from top to bottom. The red and the green areas represent the atmospheric and diffuse neutrino ``background'' considered in this work in the region of 1$^{\circ}$ around M87 position; while blue and black dotted lines represent the neutrino signal produced by pp and p$\gamma$ interactions respectively.}\label{nu-tot}
\end{figure}

\begin{figure}
\centering
\includegraphics[width=0.7\textwidth]{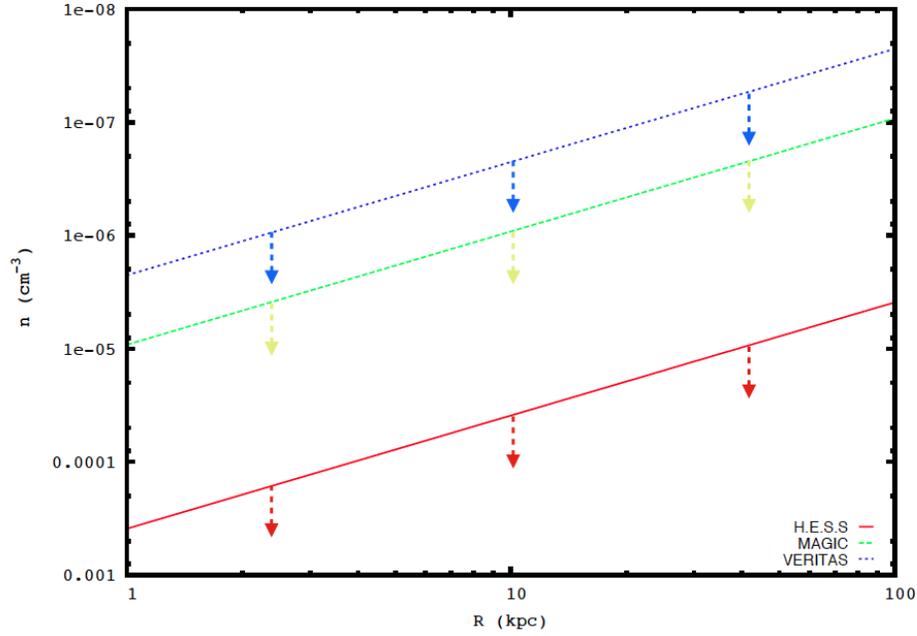}\\
\caption{Lower limit of thermal particle density as a function of  Lobes distance considering the pp model and the observations made by TA experiment in 5 years of observation. Less than one UHECR event possibly related with M87 position has been considered for this calculation.}\label{invcom}
\end{figure}

\begin{figure}
\centering
\includegraphics[width=0.8\textwidth]{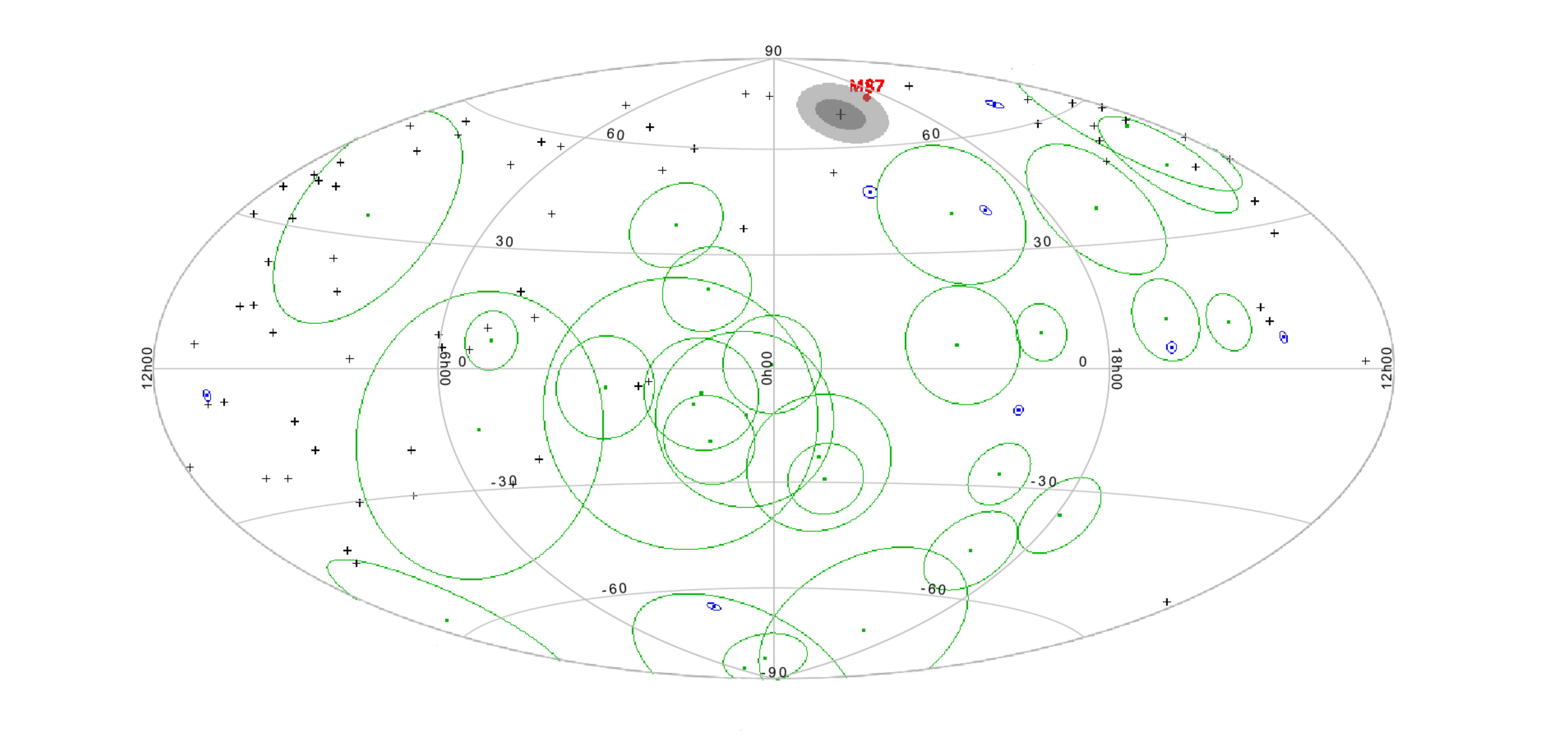}\\
\caption{Skymap in galactic coordinates: with blue points are reported the neutrino track events reported by IceCube experiment and the respective errors (blue circles), with green points and green circles the neutrino shower events with the respectively directional errors represented by the green circles. With the black crosses are reported the cosmic ray events reconstructed by TA experiment above 57 EeV in 5 years. For the closest UHECR event to the M87 position we reported two grey contour region related to the possible deviation from the original direction due to the galactic and extragalactic magnetic fields, the inner one represent a deviation from the original direction of $5^{\circ}$ while the surrounding one represent a deviation of $10^{\circ}$.}\label{Skymap}
\end{figure}

Evidence of hadronic processes occurring in the jet and in lobes of M87 radiogalaxy has been analyzed at the beginning of this work. A first hint of possible hadronic components in the gamma-ray 
emission of M87 has been obtained with the global spectral analysis of this object. The entire SED of this radiogalaxy was fit by Fermi Collaboration in 2009 with a SSC one-zone model 
\citep{2009ApJ...707...55A}. In this work we extend the spectral analysis to the VHE regime  with gamma-ray campaigns reported by H.E.S.S. with data collected in 2005-2007 \citep{2008ICRC....3..937B}, by VERITAS with data collected in 2008 \citep{2010ApJ...716..819A} and by MAGIC with data collected in 2005-2007 \citep{2012A&A...544A..96A}. In particular, we fitted the Fermi-LAT 
data obtaining the same spectral index reported by the Fermi collaboration and we extended the best fit power law up to the higher energies (see fig. \ref{fit}). This analysis pointed out the presence of 
extra components in addition to the single one-zone SSC emission, possibly of hadronic origin. A further evidence of the hadronic origin of these extra components comes from the study of the 
correlation of these gamma-ray data with the simultaneous X-ray data in the 2-10 keV energy range collected by RXTE/ASM (http://xte.mit.edu/ASM$\_$lc.html). The analysis reported in fig. \ref{event-display} comprises VERITAS nightly averaged fluxes above 250 GeV, MAGIC fluxes above 100 GeV averaged on different time scales and H.E.S.S nightly averaged fluxes above 730 GeV. Since for 
VERITAS data the exact duration of the nightly observation is unknown, we assumed that the gamma-ray fluxes are constant for a whole day and we combined them with RXTE/ASM daily data taken 
on the same observation days. In the case of MAGIC and H.E.S.S. data, we rebinned the RXTE/ASM daily and dwell\footnote{it is referred to a time bin of 90 seconds} light curves respectively, in 
order to have X-ray fluxes averaged on the same time intervals as MAGIC and H.E.S.S. 
Although the data sets can be fitted with a straight line (see fig. \ref{event-display}), all of them are weakly correlated, with low Spearman correlations coefficients (see Table 1). 
\begin{table}[ht!]
\centering
\begin{tabular}{|c|c|c|c|}
\hline
$\gamma$-experiment & Correlation coefficient & Best fit slope & Reduced $\chi^2$ \\ [0.5ex]
\hline
VERITAS & 0.01 &  (1.48 $\pm$ 1.58) $\times10^{-12}$ & 1.81 \\ [0.5ex]
MAGIC & 0.31 &  (1.79 $\pm$ 1.23)$\times10^{-12}$ & 0.47 \\ [0.5ex]
H.E.S.S. & -0.16 & (-3.52 $\pm$ 2.31)$\times10^{-13}$ &  0.60\\ [0.5ex]
\hline
\end{tabular}
\begin{center}
\scriptsize{\textbf{Table 1: Pearson correlation coefficient, slope and reduced $\chi^2$ of the best fitting straight line for the correlations}}.
\end{center}
\end{table}
Considering the reported evidences of hadronic emission components at VHE, we proceeded to describe the gamma-ray spectra of M87 as $\pi^0$ decay products occurring in p$\gamma$ and pp 
interactions. In the p$\gamma$ interaction model,  we assumed that these neutral pions are produced in the interaction of  Fermi-accelerated protons with the synchrotron self Compton (SSC) photons 
at $\sim$2 MeV encountered in the emission region. The gamma-ray spectrum generated by this process (eq. \ref{pgammam}) depends on the proton spectrum parameters (through $A_p$ and $
\alpha$), the comoving dissipation radius $r_d$,  the break energy of target photons $\epsilon_{\gamma,b}$ and the Doppler factor $\delta_{D}$. We fitted the observed spectra (see fig. \ref{fit2}) 
with the following function (see also the section 2) where $[A]$ and $[B]$ are the free parameters.
%
{\small
\bary
\label{pgammafit}
\left(\epsilon^2\,\frac{dN}{d\epsilon}\right)_{\pi^0,\gamma}= [A]  \cases{
\left(\frac{\epsilon_{\pi^0,\gamma,c}}{\epsilon_{0}}\right)^{-1} \left(\frac{\epsilon_{\pi^0,\gamma}}{\epsilon_{0}}\right)^{-[B]+3}          &  $ \epsilon_{\pi^0,\gamma} < \epsilon{\pi^0,\gamma,c}$\cr
\left(\frac{\epsilon_{\pi^0,\gamma}}{\epsilon_{0}}\right)^{-[B]+2}                                                                                        &   $\epsilon_{\pi^0,\gamma,c} < \epsilon_{\pi^0,\gamma}$\,,\cr
}
\eary
}
\begin{center}\renewcommand{\arraystretch}{0.6}\addtolength{\tabcolsep}{-1pt}
\begin{tabular}{ l c c c c c}
 \hline \hline
 \scriptsize{} & \scriptsize{Parameter} &\scriptsize{Symbol} & \scriptsize{H.E.S.S.} & \scriptsize{MAGIC} & \scriptsize{VERITAS}  \\
 \hline
\hline
\scriptsize{Proportionality constant} ($10^{-13}\,{\rm TeV/cm^2/s}$) &\scriptsize{[A]}  & \scriptsize{$ A_{p\gamma,\gamma} $}  &  \scriptsize{ $13.3\pm 0.096$} &  \scriptsize{ $3.38\pm 0.431$} &
\scriptsize{ $5.39\pm 0.94$}\\
\scriptsize{Power index}                    &\scriptsize{[B]}  & \scriptsize{$ \alpha $}  &  \scriptsize{ $2.28\pm 0.052$} &  \scriptsize{ $2.97\pm 0.121$} & \scriptsize{ $2.70\pm 0.23$} \\
\scriptsize{Chi-square/d.o.f}                                                             & & \scriptsize{$ \chi^2/{\rm d.o.f}$}  &  \scriptsize{ $14.62/7$} &  \scriptsize{ $12.59/4$} & \scriptsize{ $4.794/4$} \\
 \hline
\end{tabular}
\end{center}
\begin{center}
\scriptsize{\textbf{Table 2: the best fit of the set of p$\gamma$ interaction obtained after fitting the VHE spectrum.}}\\
\end{center}
In the pp interaction model,  we have considered  accelerated protons described by  the simple power law (eq. \ref{prot_esp}) which could be accelerated in the lobes and furthermore interact with  
thermal particles present there. The spectrum generated by this process (eq. \ref{pp}) depends on  the proton spectrum parameters (through $A_p$ and $\alpha$), the proton luminosity (through $A_p
$), number density of thermal particles and size of the lobes. We fitted the observed spectra (see fig. \ref{fit2}) with the following function (see also the section 2) with the free parameters $[A]$ and $[B]
$.
\begin{equation}
\label{pp}
\left(\epsilon^{2}_\gamma\, \frac{dN_\gamma}{d\epsilon_\gamma}\right)^{obs}_{\pi^0}= [A]\,\left(\frac{\epsilon_{\gamma,\pi^0}}{{\rm GeV}}\right)^{2-[B]},
\end{equation}
\begin{center}\renewcommand{\arraystretch}{0.6}\addtolength{\tabcolsep}{-1pt}
\begin{tabular}{ l c c c c c}
 \hline \hline
 \scriptsize{} & \scriptsize{Parameter} &\scriptsize{Symbol} & \scriptsize{H.E.S.S.} & \scriptsize{MAGIC} & \scriptsize{VERITAS}  \\
 \hline
\hline
\scriptsize{Proportionality constant} ($10^{-13}\,{\rm TeV/cm^2/s}$) &\scriptsize{[A]}  & \scriptsize{$ A_{pp,\gamma} $}  &  \scriptsize{ $12.0\pm 0.08$} &  \scriptsize{ $4.00\pm 0.43$} &\scriptsize{ $5.11\pm 0.89$}\\
\scriptsize{Power index}                    &\scriptsize{[B]}  & \scriptsize{$ \alpha $}  &  \scriptsize{ $2.22\pm 0.05$} &  \scriptsize{ $2.33\pm 0.12$} & \scriptsize{ $2.48\pm 0.20$} \\
\scriptsize{Chi-square/d.o.f.}                                                             & & \scriptsize{$ \chi^2/{\rm d.o.f.}$}  &  \scriptsize{ $13.93/7$} &  \scriptsize{ $7.28/4$} & \scriptsize{ $3.60/4$} \\
 \hline
\end{tabular}
\end{center}
\begin{center}
\scriptsize{\textbf{Table 3: the best fit of the set of pp interaction obtained  after fitting the VHE spectrum.}}\\
\end{center}
The parameters obtained with the explained fits (Tables 2 and 3) were used in this work to obtain the expected neutrino spectral parameters ($A_{(pp,\nu/p\gamma,\nu)}$, $\alpha$). Being M87 a 
close source, no extragalactic background light (EBL) absorption models \citep{2008IJMPD..17.1515R} were considered relating $A_{(pp,\gamma/p\gamma,\gamma)}$ to $A_{(pp,\nu/p\gamma,\nu)}$. 
Convoluting the obtained parameters with the effective volume ($V_{eff}$) of a hypothetical Km$^{3}$ neutrino telescope located in the northern hemisphere it was possible to obtain the 
expected neutrino events. M87 position was considered in the Monte Carlo simulation as a neutrino emitters with no cut-off for the neutrino spectra in the sensible energy range (10 GeV-10 PeV). For 
the simulated neutrino telescope we calculated also the atmospheric and cosmic neutrinos expected from the portion of the sky inside a cone centered in M87 and having an open angle of $1^{\circ}$. 
The cosmic diffuse neutrinos signal has been discussed by Waxman and Bahcall  \citep{bah01, wax98} and the upper limit for this flux is $E^{2}_{\nu}\,d\Phi/dE_{\nu}<2\times 10^{-8}GeVcm^{-2}
s^{-1}sr^{-1}$ \citep{1999PhRvD..59b3002W}. The atmospheric neutrino flux is well described by the Bartol model \citep{bar04,bar06}  for the range of energy considered in this analysis. The 
atmospheric muon ``background'' was precluded from this analysis due to earth filtration and to the softener spectrum with respect to the neutrino signal and ``backgrounds'' considered. The signal to 
noise ratio was obtained taking into account one year of observations for the considered neutrino telescope. No visible excess of M87 neutrino signal was encountered (see fig.\ref{nu-tot}) for the two 
hadronic models (pp and p$\gamma$) applied to the considered spectra (see fig.\ref{fit2}). Only neutrino flux obtained with the hadronic fit (pp) of H.E.S.S. data has the possibility to be detected in 
several years of observation with a global neutrino network\footnote{\vspace{-1cm} Global network of neutrino telescopes (IceCube, KM3NeT, ANTARES and Baikal); a infrastructure of several Km$^{3}$ and a complete coverage of the sky.} (GNN). This future scenario will be possible considering the declination of M87 and the favorable cross 
correlation between IceCube and a future northern Km$^{3}$ neutrino telescope. Linking the fit obtained for VHE gamma-ray with the accelerating proton spectra we finally estimate the UHECRs 
luminosity of  M87. In accordance with the proposed model,  we took into account a proton flux constrained by the pp interactions occurring in the lobes of M87 and set upper limits on the number of 
UHECRs in agreement  with the 5 years of observations performed by TA experiment. Taking into account the uncertainties on UHECRs arrival direction explained in section 4 and the numerical 
approach \citep{2010ApJ...710.1422R} available for extragalactic sources, we considered two contour regions of 5$^{\circ}$ and 10$^{\circ}$ around the only UHECR possibly related to M87 position 
(as shown in fig. \ref{Skymap}). With the assumption of less than one event detected in 5 years and the obtained opacity on UHE proton flux produced by pp interaction we calculated the lower limits 
of thermal proton density for different sizes of the M87, as shown in fig. \ref{invcom}. From this figure one can see that for the  lobe size $1\, {\rm kpc}<\,{\rm R}<100 \,{\rm kpc}$ the expected thermal 
proton density lies in the range $10^{-8}\,{\rm cm}^{-3}<\,{\rm n_{p}}<10^{-3}\,{\rm cm}^{-3}$. 

\section{Conclusions}
The radiogalaxy M87 was observed in the last decade at VHEs by several IACTs. The nature of the emission in this energy range is still under debate.
In this work we extended previous analysis of M87 SED up to VHEs with data collected in the last decade by MAGIC, VERITAS and H.E.S.S. We showed that the SSC one-zone scenario previously 
published by the Fermi collaboration cannot be used to describe the VHE part of the SED within the same leptonic model, therefore opening the possibility for extra hadronic components. Further 
evidences for a non-leptonic origin of the selected gamma-ray data come from the correlation of gamma-ray fluxes with simultaneous X-ray data collected by RXTE/ASM. The 
samples considered in these two different energy ranges have been found to be weakly correlated, with very low Spearman correlation coefficients. Given these results we investigated possible 
hadronic scenarios to explain the VHE emission of M87. We fitted the selected gamma-ray spectra considering proton-photon and proton-proton interactions occurring respectively in the jet and in the 
lobes of M87. For the first hadronic interaction we considered as target photons those ones at the second SSC peak, while for the second hadronic interaction we considered as a target the thermal 
protons present in the giant lobes. We verified that the samples of data collected by MAGIC, VERITAS and H.E.S.S. were well fitted by the two hadronic models and we used the best fit parameters to 
estimate the spectral neutrino counterpart. The found neutrino spectra have been convoluted with the effective volume of a hypothetical northern hemisphere Km$^{3}$ neutrino telescope, 
through Monte Carlo simulations. We obtained the signal to noise ratio for the circular region of one square degree around the position of M87. None of the selected neutrino spectra gave us a 
considerable neutrino signal excess from M87 respect to the atmospheric and the diffuse extragalactic neutrino event rates. These results have been found in accordance with what has been 
observed in four years of IceCube data-taking. However, the neutrino spectrum obtained from the H.E.S.S. data, showed a small excess of visible signal above 100 TeV that would be eventually 
visible in several years of operation of a future GNN infrastructure. Within the proposed hadronic scenario we linked the studied gamma-ray spectra with the luminosity of  accelerating protons in the 
jet and in the lobes of M87. Considering as upper limit what has been observed up to now by TA experiment we were able to put some constrains on the density of M87 lobes for different sizes. In 
particular we found that the density ranges from $10^{-8}\,{\rm cm}^{-3}$ to $10^{-3}\,{\rm cm}^{-3}$ for a lobe size in the range $1\, {\rm kpc}<\,{\rm R}<100 \,{\rm kpc}$.

\section*{Acknowledgements}
We would like to thank Michelle Hui, Matthias Beilicke, Alba Fernandez, Fabrizio Tavecchio and Karsten Berger for sharing with us the data samples used in their publications about the M87 TeV emission. Also we thank Francis Halzen, William Lee, Antonio Stamerra and Steven Neil Shore for useful discussions  and the TOPCAT team for the useful sky-map tools. This work was supported by Luc Binette scholarship and the projects IG100414, Conacyt 101958 and PAPIIT IN-108713.
%


%
%
%
%
%
\clearpage
%
%
%
%
%

\end{document}